\documentstyle[12pt]{article}

\begin{document}
{\sf \begin{center} \noindent {\Large\bf Mean-field helicity in random ${\alpha}^{2}$-dynamo twisted flows}\\[1mm]

by \\[0.1cm]

{\sl  L.C. Garcia de Andrade}\\Departamento de F\'{\i}sica
Te\'orica -- IF -- Universidade do Estado do Rio de Janeiro-UERJ\\[-3mm]
Rua S\~ao Francisco Xavier, 524\\[-3mm]
Cep 20550-003, Maracan\~a, Rio de Janeiro, RJ, Brasil\\[-3mm]
Electronic mail address: garcia@dft.if.uerj.br\\[-3mm]
\vspace{0.3cm} {\bf Abstract}
\end{center}
\paragraph*{}
 Recent results by Shukurov, Stepanov and Sokoloff [\textbf{Phys. Rev. E 78,025301,(2008)}] on dynamo action in Moebius strip flows, indicate that strong magnetic fields appear in regions where, electric current and flow kinetic helicities have the same sign. Here, an analytical version of their numerical result is obtained by a simpler method of considering the laminar non-turbulent limit, of a twisted Riemannian thin flux tube. It is shown that the magnetic field is amplified, when electric current helicity and Riemann curvature are both negative. Thus spaces of positive and negative Riemannian curvatures seems to support dynamo action inside the torus, and not only negative Riemannian curvature surfaces as happens in 2D dynamos. New features appear in ${\alpha}^{2}$-dynamo twisted flow, using the approximation of thin tubes flux tubes. These solutions are obtained in the resonant profile of the toroidal and poloidal frequencies modes of the dynamo force-free flow. To a certain extent our solution seems to extend Shukurov, Stepanov and Sokoloff, solution to more general twisted flows than their Moebius strip flow. From Perm liquid sodium torus dynamo experimental data, torsion of the geometrical axis of the twisted torus is computed. For an external radius of the torus of $R\approx{8.0{\times}10^{-2}}m$ yields a strong torsion of the order of $10 m^{-1}$. Knowledge of the topological structure of flows helps to design future dynamo experiments. Negative Riemannian curvature flows may appear in solar inflexionary flux tubes, where internal curvature is positive while external curvature is negative, and therefore the scalar Gaussian or two-dimensional curvature is negative. Simple Arnold-Beltrami-Childress (ABC) dynamo flows are obtained. Mean-field ABC dynamos were developed by Kleeorin et al [Phys Rev \textbf{E79},046302,(2009)] for anisotropic turbulence.{\bf PACS numbers:\hfill\parbox[t]{13.5cm}{47.65.Md,Hw.47.65.-d}}}
\newpage \section{Introduction}
 Kinematic screw (helical) dynamo action
 \cite{1} in the flow of a cylindrical periodic tube conducting wall, has been
 investigated recently by Dobler, Frick and Stepanov \cite{2}. Spectral analysis in the steady regime, have been used to investigate dynamo action in that device. Recently  creation of helical turbulence by dynamo flows in the Perm dynamo liquid sodium experiment \cite{3,4,5}, has been done by using a rotating
 torus device, where turbulence is created by a sudden braking of the rotating torus by an external mechanism. Hydrodynamical turbulence can be simulated with water instead of the usual Gallium or Sodium liquid metals. As a consequence, dynamo action in the turbulent flow, is supported inside the metal torus. This Riemann flat toroidal space of the dynamo experiment, is embbeded in the Euclidean three-dimensional $\textbf{E}^{3}$ space of laboratory where a dynamo action is obtained by stretching magnetic lines by a dynamo flow. This mechanism of stretching the magnetic field lines to generate dynamo action \cite{1} first appears in the literature in a paper by Arnold, Zeldovich,
 Ruzmaikin and Sokoloff \cite{6} in the case of a uniform stretching. This paper represents the first exact analytical solution of fast dynamos. More recently, two examples of dynamo action in Riemannian space \cite{7,8} have been presented.
 The first example, was a stretch-twist and fold fast dynamo action \cite{9} in
 conformal Riemannian manifolds. The other example, is an application
 of the anti-fast-dynamo theorem, previously proposed by Vishik \cite{10}, to the plasma
 devices has been investigated. Yet more recently, another example of the use of curvilinear and twisted coordinates systems has been throughly studied by Shukurov, Stepanov and Sokoloff \cite{11}, have been proposed a Moebius strip flow, which supports
 dynamo action. In their paper quite interesting physical results on dynamo theory have been obtained such as the enhancement and amplification of dynamo action in places where the kinetic flow and magnetic helicities, respectively given by ${\alpha}=-\frac{{\tau}^{*}}{2l}\textbf{v}.{\nabla}{\times}\textbf{v}$ and ${\gamma}=\textbf{B}.{\nabla}{\times}\textbf{B}$ have the same sign. In this paper a analytical solution of the induction equation is found for isotropic turbulence, which shows that curvature of the twisted surface of the flow, plays an important role in this aspect of the experiment. Besides the Riemann curvature tensor, ${R^{1}}_{212}$ given in two dimensions by the Gaussian curvature K which is given by the product of the principal directions of torus geometry in this experiment. In the torus case the overall Gaussian curvature vanishes, nevertheless the lower internal part of the torus possesses a negative curvature while the external part possesses a positive curvature. The twist of the dynamo flow shows that, when Riemannian curvature is negative electric current and kinetic flow helicities coincide and are positive, while a change in helicities sign implies necessary a change in the curvature sign. When Riemann curvature vanishes the magnetic helicity vanishes. So one must conclude that in this kind of experiment helicity and twist associated with Riemann curvature are fundamental ingredients for dynamo action. Though the magnetic axis is torsionless, the experiment can be modified in the form of a stellarator in plasma physics to accommodate torsion effects on dynamo flow. Another interesting feature of the Perm torus liquid sodium dynamo device, is that the dynamo oscillates in space. By making use of a ${\alpha}^{2}$-dynamo here, in the limit of non-turbulent laminar flows, one is able to show that oscillatory modes for the ${\alpha}$-helicity is obtained. Besides reproducing many carachteristics of Perm experiment the the Moebius dynamo flow proposal, here one assumes that the magnetic Reynolds $R_{m}$ is as small as $R_{m}\approx{16}$. Numerical simulations of Perm dynamo for twisted Moebius strip flow, indicate that when the diffusivity vanishes slow dynamo is obtained. This paper is organized as follows: In section 2 the perturbed equation is solved and the equation between torsion of the dynamo flow and the ratio between toroidal and poloidal components of the flow is deduced. In this same section the solution for oscillating ${\alpha}^{2}$-dynamo is given. In this same section isotropic turbulence, where kinetic helicity vanishes is considered to obtain the equation for the perturbed field. Section 3 shows that mean field ABC flows is induced by a steady perturbation in the case of isotropic turbulence. In this case the dynamo is not kinematic since the initial field depends upon the speed of the perturbed flow. A true back-reaction non-linear mechanism. More involved anisotropic turbulent ABC dynamo flows have been found recently by Kleeorin et al \cite{12}. Section 4 presents discussions and conclusions.
 \newpage
\section{Random dynamo twisted flows in torus devices}
Let us now considering the equation for the pseudo-isotropic turbulent dynamo where a random magnetic flow field $<\textbf{B}^{0}>$ is perturbed according to the rule
\begin{equation}
\textbf{B}= <\textbf{B}^{0}>+{\textbf{B}_{1}}\label{1}
\end{equation}
where ${\textbf{B}_{1}}$ is the magnetic field steady
perturbation. Here, $<\textbf{B}^{0}>$ is the random applied field.
By substitution of this expression into the self-induction equation
\begin{equation}
{\partial}_{t}\textbf{B}=
{\nabla}{\times}(\textbf{v}{\times}\textbf{B})+{\eta}{\Delta}\textbf{B}\label{2}
\end{equation}
yields the expression
\begin{equation}
{\partial}_{t}<\textbf{B}^{0}>={\nabla}{\times}(\textbf{v}{\times}\textbf{B}_{1}-<\textbf{v}{\times}\textbf{B}_{1}>)
+{\eta}{\Delta}<\textbf{B}^{0}>\label{3}
\end{equation}
is obtained. Here the mean field electromotive force is given by
\begin{equation}
{\cal{E}}=<\textbf{v}{\times}\textbf{B}_{1}>\label{4}
\end{equation}
which in turn is given by
\begin{equation}
{\cal{E}}={\alpha}<\textbf{B}^{0}>-{\beta}{\nabla}{\times}{<\textbf{B}^{0}>}\label{5}
\end{equation}
where ${\alpha}=-\frac{{\tau}^{*}}{2l}[\textbf{v}.{\nabla}{\times}\textbf{v}]$, is the kinetic helicity and the turbulent diffusivity is given by ${\beta}$. Here ${\eta}$ is the magnetic diffusivity. According to Raedler and Krause \cite{12}, the pseudo-isotropic turbulence is given by the non-vanishing of ${\alpha}$ while ${\beta}$ vanishes. While in the true isotropic turbulence, ${\alpha}$ vanishes. In this section one shall make the case of pseudo-isotropic turbulent twisted flow in the laminar limit where this turbulence vanishes as well as a fast dynamo test is given for ${\eta}$ vanishing. It is shown in this section that for the pseudo-isotropic turbulence a fast dynamo solution may be obtained while as shown in the next section slow dynamos are found in the isotropic case, called by Mestel \cite{13} the genuine dynamo. In the Shukurov et al paper, it is shown that numerical simulations of another kind of twisted dynamos given by Moebius strip dynamo flow in a toroidal channel, leads to a slow dynamo. Thus by considering the twisted flow in magnetic flux tube Riemann metric
\begin{equation}
dl^{2}= dr^{2}+r^{2}d{{\theta}_{R}}^{2}+K^{2}(r,s)ds^{2} \label{6}
\end{equation}
By taking $K(r,s)=(1-{\kappa}_{1}r\cos{\theta}):=1$ , where  ${\kappa}_{1}$
is the Frenet external curvature and the twist transformation angle
\begin{equation}
{\theta}(s):={\theta}_{R}-\int{{\tau}(s)ds} \label{7}
\end{equation}
one obtains, the Riemannian line element of the thin flux tube, as
\begin{equation}
dl^{2}= dr^{2}+r^{2}d{{\theta}_{R}}^{2}+ds^{2} \label{8}
\end{equation}
Throughout the paper one shall consider the thin flux tubes approximation, which is suitable for solar dynamo flux tubes. Though the solar dynamos in general consider the ${\alpha}{\Omega}$-dynamo including differential rotation ${\Omega}$ to stretching the tube instead of the ${\alpha}^{2}$-dynamo consider here. Of course, ${\alpha}^{2}$-dynamos can be used in solar turbulent flows. Actually laminar limit, one shall address here, is in the upstream part of the solar tube, while turbulent plasma is in the downward part of the tube. Here the tube seed magnetic field is given by
\begin{equation}
<\textbf{B}^{0}>=e^{{\gamma}t}[B^{s}\textbf{t}+B^{\theta}\textbf{e}_{\theta}]\label{9}
\end{equation}
where care must be exercised in the contravariant components of the magnetic field components. Now let us split the self-induction equation above according to perturbation scheme, where $$|\textbf{B}_{1}|<<<|<\textbf{B}^{0}>|$$. The following basis
\begin{equation}
{\partial}_{s}\textbf{e}_{\theta}=-{\tau}_{0}\sin{\theta}\textbf{t}\label{10}
\end{equation}
and
\begin{equation}
\textbf{e}_{\theta}=-\sin{\theta}\textbf{n}+\cos{\theta}\textbf{b}\label{11}
\end{equation}
\begin{equation}
\textbf{e}_{r}=\cos{\theta}\textbf{n}+\sin{\theta}\textbf{b}\label{12}
\end{equation}
where $(\textbf{t},\textbf{n},\textbf{b})$ is the Frenet basis attached to the geometrical or magnetic axis of the tube, and $(\textbf{e}_{r},\textbf{e}_{\theta},\textbf{t})$ have the first two vectors attached to the tube surface. Actually in the case of a thin tube these basis are actually quite close topologically. With the aid of these equations one may solve the following perturbed equations
\begin{equation}
{\partial}_{t}<\textbf{B}^{0}>={\nabla}{\times}({\cal{E}}+{\eta}{\Delta}<\textbf{B}^{0}>)
\label{13}
\end{equation}
and
\begin{equation}
{\partial}_{t}<\textbf{B}_{1}>=
{\nabla}{\times}(\textbf{v}{\times}\textbf{B}_{1}-
[{\alpha}-{\beta}{\gamma}]{\gamma}<\textbf{B}^{0}>)+{\nabla}{\times}
(\textbf{v}{\times}<\textbf{B}^{0}>)
\label{14}
\end{equation}
where the term inside the brackets was obtained by assuming the magnetic field on the tube was a force-free field given by
\begin{equation}
{\nabla}{\times}<\textbf{B}^{0}>={\gamma}<\textbf{B}^{0}>
\label{15}
\end{equation}
and the the Laplacian ${\Delta}={\nabla}^{2}$ operator, would obey the following relation
\begin{equation}
{\nabla}^{2}<\textbf{B}^{0}>=-curl(curl<\textbf{B}^{0}>)
\label{16}
\end{equation}
Let us now solve the main equation for the initial magnetic field above. This yields the following expression
\begin{equation}
[{\lambda}-{\alpha}{\gamma}]<\textbf{B}^{0}>+
[{v}^{s}-\frac{{{\tau}_{1}}^{-1}}{r_{0}}v^{\theta}][B^{\theta}{\partial}_{s}\textbf{e}_{\theta}+B^{s}\textbf{n}{\kappa}_{0}]=
0
\label{17}
\end{equation}
where in this equation one has considered that the case of turbulent diffusion-free, and magnetic diffusion-free. A simple solution of this equation, can be obtained with the help of the divergence-free magnetic field equation
\begin{equation}
{\nabla}.<\textbf{B}_{0}>=0\label{18}
\end{equation}
which yields
\begin{equation}
\frac{1}{\sqrt{g}}[{\partial}_{i}(\sqrt{g}{B}^{i})]=0
\label{19}
\end{equation}
Since $g={r_{0}}^{2}$, where $r_{0}$ is the constant circular cross-section of the internal radius of the torus or flux tube, one obtains
\begin{equation}
{\partial}_{s}[{B}^{s}-\frac{{{\tau}_{1}}^{-1}}{r_{0}}B^{\theta}]=0
\label{20}
\end{equation}
and since the flow is also divergence-free or incompressible, one obtains
\begin{equation}
{\partial}_{s}[{v}^{s}-\frac{{{\tau}_{1}}^{-1}}{r_{0}}v^{\theta}]=0
\label{21}
\end{equation}
Thus a simple ansatz for the solution of these equations which will tremendously simplify the solution of dynamo equations are
\begin{equation}
{B}^{s}=\frac{{{\tau}_{1}}^{-1}}{r_{0}}B^{\theta}
\label{22}
\end{equation}
and
\begin{equation}
{v}^{s}=\frac{{{\tau}_{1}}^{-1}}{r_{0}}v^{\theta}\label{23}
\end{equation}
which shall give the equation for the uniform twist
\begin{equation}
\frac{{B}^{\theta}}{B^{s}}={{\tau}_{1}}{r_{0}}
\label{24}
\end{equation}
A result obtained by Ricca \cite{14} in the investigation of solar flux tubes with twist and torsion ${\tau}_{0}$. From these equations, one immeadiatly notes that equation (\ref{17}) reduces to
\begin{equation}
[{\lambda}-{\alpha}{\gamma}]<\textbf{B}^{0}>=0
\label{25}
\end{equation}
In this equation use was made of the Frenet evolution equations
\begin{equation}
\frac{d\textbf{t}}{ds}={\kappa}_{1}(s)\textbf{n}
\label{26}
\end{equation}
\begin{equation}
\frac{d\textbf{n}}{ds}=-{\kappa}_{1}(s)\textbf{t}+{\tau}_{1}\textbf{b}
\label{27}
\end{equation}
\begin{equation}
\frac{d\textbf{b}}{ds}=-{\tau}_{1}(s)\textbf{n}
\label{28}
\end{equation}
Therefore equation (\ref{25}) immeadiatly yields a solution for the random magnetic field spectrum as
\begin{equation}
{\lambda}={\alpha}{\gamma}
\label{29}
\end{equation}
which immeadiatly shows that only a slow dynamo which is a critical value for a slow dynamo, is obtained when the helicity ${\alpha}$ vanishes. On the other hand when the kinetic flow helicity is present, one obtains that signs of kinetic and magnetic helicity should be the same in order that the $lim_{{\eta}{\rightarrow}{0}}{\lambda}\ge{0}$, or a fast dynamo action appears. In the next section one shall compute the kinetic mean-field flow helicity and substitute back into the above relation to obtain the growth rate and dynamo action contraints.\newpage
\section{${\alpha}^{2}$-dynamo and ABC flows in Riemannian twisted space}
To get analytical dynamo solutions is rather difficulty and in general from dynamo experiments is only possible to design numerical experiments and sometimes is not easy to know whether the dynamo is fast or slow. In this section one finalizes the analytical solution for mean-field electrodynamics \cite{13} flow and electric current or magnetic helicities, in the case of laminar limit of diffusity turbulent coefficient ${\beta}$ vanishes along with the limit of magnetic diffusion vanishing, in order to check for the nature of the dynamo, allowing us to decide if the dynamo is fast or slow. To compute the magnetic helicity, let us use the expression
\begin{equation}
{\nabla}{\times}\textbf{B}=\frac{1}{\sqrt{g}}{\epsilon}^{ijk}[{\partial}_{j}B_{k}]\textbf{e}_{i}
\label{30}
\end{equation}
which yields
\begin{equation}
{\nabla}{\times}\textbf{B}=\frac{1}{r_{0}}[{\partial}_{\theta}B_{s}-{\partial}_{s}B_{\theta}]\textbf{e}_{r}
\label{31}
\end{equation}
From this expression is easy to show that for the torus surface the magnetic helicity vanishes if one assumes that the radial component of the magnetic field vanishes. This can be shown by simply considering that in this case $<\textbf{B}_{0}>.\textbf{e}_{r}=0$. Actually for a constant internal radius $r_{0}$ the magnetic helicity vanishes as can be easily seen from these expressions.
This means that the magnetic random field remains tangent to the torus surface and other ergodic internal surfaces. Thus to keep the effects of the magnetic helicity one might have a more general expression for the curl of the magnetic field might depend on the variation of radial directions. This is
\begin{equation}
{\nabla}{\times}\textbf{B}={{\kappa}_{0}}[({\partial}_{\theta}B_{s}-{\partial}_{s}B_{\theta})
\textbf{e}_{r}-{\partial}_{r}B_{\theta}\textbf{t}-{\partial}_{r}B_{s}\textbf{e}_{\theta}]
\label{32}
\end{equation}
Thus one may easily seen that the presence of magnetic helicity, is here guaranteed even if the radial component of the magnetic field is absent. To compute the magnetic helicity, one notes that now the expression for the equations for magnetic helicity becomes
\begin{equation}
{\partial}_{\theta}B_{s}-{\partial}_{s}B_{\theta}=0
\label{33}
\end{equation}
\begin{equation}
{{\kappa}_{0}}^{-1}{\partial}_{r}B_{s}=-{\gamma}B_{\theta}
\label{34}
\end{equation}
and
\begin{equation}
{{\kappa}_{0}}{\partial}_{r}{B}_{\theta}=-{\gamma}{B}_{s}
\label{35}
\end{equation}
where ${\kappa}_{0}=\frac{1}{r_{0}}$ is the internal curvature. One must note from this expression that the growth of the magnetic energy density grows in the radial direction, amplifying the magnetic fields and therefore enhancing dynamo action, when the external torus curvature ${\kappa}_{1}$ and the magnetic helicity have opposite signs. Thus one has that in the twisted flows of Riemannian negative curvature, where the Riemann curvature tensor is
\begin{equation}
{R^{1}}_{212}={\kappa}_{1}{\kappa}_{0}
\label{36}
\end{equation}
the magnetic helicity and external curvature shall possesses the opposite sign. From the above expressions and the relation between partial derivative operators ${\partial}_{\theta}=-{{\tau}_{1}}^{-1}{\partial}_{s}$, one obtains the following expressions
\begin{equation}
B_{\theta}={\partial}_{\theta}{\psi}(s)=-{{\tau}_{1}(s)}^{-1}{\partial}_{s}{\psi}(s)=-{{\tau}_{1}(s)}^{-1}B_{s}\label{37}
\end{equation}
\begin{equation}
B_{\theta}=B_{0}(s)e^{{\gamma}{{\kappa}_{1}}r}
\label{38}
\end{equation}
To obtain the function ${B_{0}}(s)$ one must use the divergence-free property of the magnetic dynamo field, or in physical terms the absence of magnetic monopoles condition. Here one is forced to consider that, the external curvature ${\kappa}_{1}$ depends on the toroidal coordinate-s. Thus the expression for the toroidal field obtained is
\begin{equation}
B_{s}=(1+{{\kappa}_{1}}^{-2})e^{{\gamma}{{\kappa}_{1}}r}
\label{39}
\end{equation}
Note that from the Perm dynamo experimental data, where ${R}=0.08 m$, the ${{\kappa}_{1}}^{-2}=R^{2}\approx{6.4{\times}10^{-3}}m^{2}$ which is a extremely low value and the expression can be approximated by
\begin{equation}
B_{s}\approx{e^{{\gamma}{{\kappa}_{1}}r}}
\label{40}
\end{equation}
The poloidal field may be obtained from the above relation between the two and torsion.
The magnetic energy density can now be easily computed as
\begin{equation}
{\epsilon}=\frac{1}{8{\pi}}[{B_{\theta}}^{2}+{B_{s}}^{2}]
\label{41}
\end{equation}
which yields
\begin{equation}
{\epsilon}=\frac{1}{8{\pi}}[1+{{\kappa}_{1}}^{-1}]e^{2{\gamma}{\kappa}_{1}r}
\approx{e^{{\gamma}{\kappa}_{1}r}}
\label{42}
\end{equation}
Note that the grow of this expression depends exactly of the interaction between magnetic helicity and the external curvature of the twisted torus. Since the internal curvature of the twisted flow is ellipsoidal or circular \cite{11}, and therefore possesses a positive curvature the sign of the Riemann curvature is determined only by ${\kappa}_{1}(s)$. Therefore to have a growth in the magnetic energy density (\ref{42}) the magnetic helicity and Riemann curvature of the twisted flow must possess the same sign. Note also that in this laminar $[{\beta}=0]$ limit of the isotropic turbulence, the ABC flow shows that the equation for kinetic flow energy is similar
\begin{equation}
{k}=\frac{1}{8{\pi}}[1+{{\kappa}_{1}}^{-1}]e^{2{\gamma}{\kappa}_{1}r}
\approx{e^{{\gamma}{\kappa}_{1}r}}
\label{43}
\end{equation}
Thus to obtain an explicity form for growth rate given by (\ref{29}) the mean-field flow helicity ${\alpha}$ might be computed as
\begin{equation}
{\alpha}\approx{<e^{{\kappa}_{1}{\gamma}r}>}
\label{44}
\end{equation}
\begin{equation}
{\alpha}=-\frac{{\tau}^{*}}{l}[\sinh({\gamma}{\kappa}_{1}r)+\cosh({\gamma}{\kappa}_{1}r)]
\label{45}
\end{equation}
which yields the following growth rate
\begin{equation}
lim_{({\beta},{\eta})\rightarrow{(0,0)}}{\lambda}=-\frac{{\tau}^{*}}{l}{\gamma}[\sinh({\gamma}{\kappa}_{1}r)+\cosh({\gamma}{\kappa}_{1}r)]
\label{46}
\end{equation}
Finally let us estimated the twisted torus torsion, by making use of the Perm dynamo torus external radius ${R}=0.08 m$. This yields a torsion of the order ${\tau}_{0}\approx{10}m^{-1}$. In the next section one shall consider the the case of natural appearence of ABC flows when the
\section{Conclusions}
Recently Kleeorin et al \cite{12} have investigated the random magnetic field evolution of magnetic field based on a mean-field magnetohydrodynamics, in two distinct type of flows called Roberts flow and ABC flows. Turbulence even in experimental sodium dynamo devices such as the Perm one \cite{11}, are in general difficulty to handle in analytical form such as in the case of Roberts and ABC flows. Moebius strip dynamo action considered recently by Shukurov et al \cite{11} has been handled by numerical simulations, and so far no analytical solutions have been obtained for this difficult to solve this problem, To countour the situation and bypass it, here one studies a type of twisted flows, other than Moebius strip one called a twisted magnetic flux tube, which simulates a flow inside the Perm torus dynamo. To obtain simpler solutions, one assumes the laminar limit of isotropic turbulence. The mean-field value of the helicity is computed and the growth rate of the magnetic field is obtained. Under appropriated conditions not only slow dynamos can be obtained but also fast ones. A yet more general type of Riemannian twisted geometry called thick twisted flows in torus, can be obtained in near future. Since as shown by Schuessler \cite{15} solar flux tube dynamos can be obtained without the help of mean-field MHD, to have a laminar, non-turbulent limit of the turbulent dynamos in twisted flux tubes, as done here seems to be a good motivation for further investigation.
\newpage\section{Acknowledgements} I also am deeply indebt to Dmitry Sokoloff, Javier Burguette, Guenther Ruediger and Renzo Ricca for helpful discussions on
the subject of this paper. Financial supports from Universidade do
Estado do Rio de Janeiro (UERJ) and CNPq (Brazilian Ministry of
Science and Technology) are highly appreciated.

 \end{document}